\documentclass[prd,twocolumn,minimal,amsmath,amssymb,showpacs,floatfix,superscriptaddress,nofootinbib,preprintnumbers]{revtex4}
\usepackage{graphicx}
\usepackage{psfrag}
\usepackage{epsfig}
\usepackage{bm}
\usepackage{amsfonts}
\usepackage{float}
\usepackage{color}
\usepackage{comment}
\usepackage{amsmath,amssymb}
\def\gev{{\rm \,Ge\kern-0.125em V}}

\def\calR{{\cal R}}
\def\be{\begin{equation}}
\def\ee{\end{equation}}
\def\bea{\begin{equationarray}}
\def\eea{\end{equationarray}}

\begin{document}

\title{Scalar spectral index in the presence of Primordial Black Holes}

\author{Gaveshna Gupta }
\email{gaveshna.gupta@gmail.com} 
\affiliation{Department of Physics $\&$ Astrophysics, University of Delhi, New Delhi 110 007 India.}
\author{Ramkishor Sharma}
\email{sharmaram.du@gmail.com}
\affiliation{Department of Physics $\&$ Astrophysics, University of Delhi, New Delhi 110 007 India.}

\author{T.R. Seshadri}
\email{trs@physics.du.ac.in}
\affiliation{Department of Physics $\&$ Astrophysics, University of Delhi, New Delhi 110 007 India.}

\begin{abstract}
 
We study the possibility of reheating the universe in its early stages through the evaporation of Primordial Black Holes (PBHs) that are formed due to the collapse of the inhomogeneities that were generated during inflation. By using the current results of the baryon-photon ratio obtained from BBN and CMB observations, we impose constraints on the spectral index of perturbations on those small scales that cannot be estimated through CMB anisotropy and CMB distortions. The masses of the PBHs constrained in this study lie in the range of $10^{9}$ and $10^{11}$g, which corresponds to those PBHs whose maximal evaporation took place during the redshifts $10^{6} < z < 10^{9}$. It is shown that the upper bound on the scalar spectral index, $ n_{s}$ can be constrained for a given threshold value, $ \zeta _{\rm th}$, of the curvature perturbations for PBHs formation. Using Planck results for cosmological parameters we obtained $n_{s} < 1.309 $ for $ \zeta _{\rm th} =0.7 $ and $ n_{s} < 1.334 $ for $ \zeta _{\rm th} = 1.2 $ respectively. The density fraction that has contributed to the formation of Primordial Black Holes has also been estimated.

\end{abstract}
\pacs{ 98.80-k, 98.80.Cq, 04.70.Dy,}
\maketitle

\section{Introduction}
\noindent
The universe was to a high degree homogeneous and isotropic in its early phase, in which tiny fluctuations are believed to have been generated during the inflation \citep{Carr1994,kholpov-inf,Naselsky-inf} due to quantum processes. The fluctuations \cite{zeldovich,Carr1975} of cosmological relevance were smaller than the Hubble radius ($H^{-1}$) at the onset of inflation, exited $H^{-1}$ to  grow to super horizon scales during inflation and re-entered during the post inflationary phase. If these density perturbations are of size of the order of horizon scale and amplitude above a certain critical threshold a few regions become sufficiently compressed for gravity to overpower pressure forces and the rate of expansion and in turn cause collapse to a black hole\cite{Hawking1971}. Such black holes are being referred as Primordial Black Holes (PBHs). Carr and Hawking(1974)\cite{carr-hawk} also showed
that the PBHs would not have grown much due to accretion which implies that their masses today should be same as it was during its formation. In general, the collapsing region during the radiation dominated epoch is assumed to be spherically symmetric \cite{dorosh,bardeen}. The critical density required for a region to collapse and form a PBH strongly depends on the density profile of the overdense region.

The PBH mass at the time of its formation can be estimated to be equal to or smaller than the cosmic horizon mass.
\begin{equation}
M_{H}\approx 10^{15} \left(\dfrac{t}{10^{-23}} \right) g \,.
\end{equation}
Unlike stellar mass black holes, which have a minimum mass criterion a range of different masses are possible in the case of PBHs.
Those for which the mass was less than $10^{15}$g at the time of their formation must have evaporated away by the present epoch by virtue of Hawking radiation\cite{hawking-nature} i.e, black body radiation being emitted from a black hole at a temperature which is inversely proportional to its mass. On the other hand, PBHs whose mass $M > 10^{15}$g could survive in the present universe.

In principle, the formation of PBHs could also take place due to several other mechanisms\cite{Carr2009,Hawking-1989,Garriga,Caldwell,Gibbon-1998} like topological defects and bubble collisions\cite{crawford-bubble,Hawking-bubble,kodama,La}. In the observable universe, the PBHs abundance can constrain the primordial power spectrum,
and hence models of inflation, on scales much smaller than those that are accessible via observations of the CMB and LSS (which provide strong constraints on the primordial power spectrum for scales between k $\sim 10^{-3}-1$ Mpc$^{-1}$). Thus, PBHs can act as a significant probe to study the earliest phase of the universe complimenting the information from CMB and LSS analysis.

Our work in this article is organised as
follows.  In section \ref{reheating} we describe how the evaporating PBHs can inject energy into the photon-baryon fluid in the early stages of our universe. In section \ref{estimate-energy} we describe how we have estimated the energy injected into the background fluid via Hawking radiation. In section \ref{pbh-abund} we describe the constraints on $\beta(M)$ which reflects the abundance of regions of mass M that collapses to form PBHs. In section \ref{constraints} we try to obtain the upper limit on the spectral index of perturbations with a power law type power spectrum and contrast it in light of WMAP data and Planck data. We end with our conclusions in section \ref{concl}.

\section{Reheating by PBH\MakeLowercase{s}}\label{reheating}
In the early universe, the PBHs inject energy (produced by the process of Hawking radiation) into photon-baryon fluid. If this energy injection happens before the CMB distortion era ($ z\gtrsim z_{\mu} = 2\times 10^{6}$) the fluid can achieve black-body spectrum by photon-electron interaction via double Compton scatterings and Bremsstrahlung \cite{hecklerprl,hecklerprd,Page:2007yr} by the epoch of recombination. This is so because the time scales of the interaction for these processes is smaller than the cosmological time scale. However, these processes (double Compton process and Bremsstrahlung) result in the increase of photon number, while the total number of baryons remains unchanged. Consequently, a baryon-photon ratio $ \eta \equiv n_{b}/n_{\gamma} $ decreases due to this energy pumped due to evaporating PBHs.

% Thus, the injected energy from the PBHs would not be observed in the form of CMB distortions on small scales. (Note that these distortions are typically characterized by two parameters namely, the chemical potential $ \mu $ and the $ y $ Compton parameter. \cite{})

In the $z>z_{\mu}$ era, our focus, in particular, lies in exploring the thermal history of the universe between $z_{\mu}=2\times 10^{6}$ and $z_{i}=1\times 10^{9}$. This redshift range corresponds to energy injections of those PBHs whose mass range lies between $10^{9}$ and $10^{11}$g at the time of their formation.

From an observational point of view, both BBN and CMB provide constraints on $ \eta $ at their respective epochs. If a process pumps energy in a redshift interval $z$, $z+dz$ we can define the fractional density excess as 
\begin{equation}\label{energy-inj}
\dfrac{\Delta \rho_{\gamma}}{\rho_{\gamma}} = \int dz \dfrac{1}{\rho_{\gamma}(z)} \dfrac{dQ}{dz}\,,
\end{equation} 
where $\rho_{\gamma}(z)$ is the background energy density of photons which scales as $(1+z)^{4}$.

The ratio of the number of baryons per photon at recombination ($\eta_{CMB}$) to that at BBN ($\eta_{BBN}$) is related to the fractional density excess of the energy pumped between the epoch of BBN and recombination as \cite{Nakama:2014vla}
\begin{equation}
\dfrac{\eta_{CMB}}{\eta_{BBN}} = \left(1-\dfrac{3}{4}\dfrac{\Delta \rho_{\gamma}}{\rho_{\gamma}}\right) \,.
\end{equation} 

Nakama et al.  have used the observationally inferred values of $ \eta_{CMB}$ and $\eta_{BBN}$ as 
$(6.11 \pm 0.08)\times 10^{-10}$ and $(6.19 \pm 0.21)\times 10^{-10} $ respectively \cite{Ade:2015xua, Nollett:2013pwa}. Using these values, they obtained an upper bound on the density fraction of the injected energy as 
\begin{equation}\label{density-bound}
\dfrac{\Delta \rho_{\gamma}}{\rho_{\gamma}} < 7.71 \times 10^{-2} .
\end{equation}

In our study, we calculate the fractional density excess for PBHs evaporation. This fractional density excess depends on $n_s$. Further, using Eq.\eqref{density-bound} we place a bound on $n_s$.

For a certain region to collapse into a black hole for a flat Friedmann background, it was found that 
$\delta (\equiv \frac{\Delta M}{M}$, the fractional mass difference within the initial horizon) should be of the order of $1/3$ \cite{Carr1975}.
 
Shibata and Sasaki \cite{Shibata:1999zs} had obtained, using numerical simulations, the threshold value of density fluctuations required for the overdense regions to collapse into PBHs in the horizon crossing epoch. We use these results in our work.

The gauge-invariant curvature perturbation $\zeta $ on the uniform-density hyper surface is defined as \cite{Wands:2000dp}
\begin{equation}
\zeta ={\cal R} - H {\delta \rho \over \dot \rho}\,,
\label{zeta-def}
\end{equation}
where $\rho$ and $\delta \rho$ are the background density and the perturbed
density, respectively, ${\cal R}$ is the curvature perturbation and dot (.) represents
the time derivative. Green et al.\cite{Green:2004wb} derived the relation between the threshold of density perturbations and the threshold value of $\zeta$ ($\zeta_{th}$). Corresponding to a threshold density perturbation of 0.5 and 0.3  the $\zeta_{th}$ turns out to be 1.2 and 0.7 respectively \cite{Green:2004wb}.

In our analysis, we assume a power-law primordial power spectrum, ${\cal P}_{{\cal R}} ={{\cal
R}_c} (k/k_0)^{n_{s}-1} $. We have also used ${{\cal R}_c} = (24.0 \pm 1.2) \times 10^{-10}$ at the scale $k_0 =0.002 {\rm Mpc}^{-1}$ from WMAP \cite{Spergel:2006hy} as well as ${{\cal R}_c} = (21.39 \pm 0.063) \times 10^{-10}$ at the scale $k_0 = 0.05 {\rm Mpc}^{-1}$ from Planck \cite{Ade:2015xua} to calculate ${\cal P}_{{\cal R}}$ in each case. When the density fields are smoothed by a Gaussian window function with comoving size $R$, the peak theory gives the comoving number density of the peaks which are higher than $\nu$ and can be expressed in the high peak limit($\nu\gg 1$) as \cite{Green:2004wb}
\begin{equation}
n (\nu, R) =
{1 \over (2 \pi)^2} {(n_{s}-1)^{3/2} \over 6^{3/2} R^3} {(\nu^2 -1)}
\exp\left(-{\nu^2 \over 2}\right)\,.
\label{density-peak}
\end{equation}
Here $\nu$ can be expressed in terms of the threshold value of the curvature perturbation $\zeta_{th}$ for the PBH formation by
\begin{equation}
\nu =
\left[{2 (k_0 R )^{n_{s}-1} \over {{\cal R}_c} \Gamma ((n_{s}-1)/2) }\right]^{1/2}
\zeta _{\rm th}\,.
\label{nu-threshold}
\end{equation}
We consider Eq. (\ref{density-peak}) to be the comoving number density
of PBHs formed from the collapse of the overdense regions with scale $R$.

The smoothing scale $R$ can be related to the PBH mass which further
depends on the initial environment around the peak as well as the
threshold value. For the sake of simplicity, we assume that PBHs with the horizon mass are produced when the regions with overdensity above the threshold value enter the horizon. The scale $R$ at this epoch is
given by
\begin{equation}
R= {1 \over a H}\,.
\label{formation-time}
\end{equation}
The comoving entropy conservation gives
\begin{equation}
g_* ^{1/3} a T= const.\,,
\label{temperature-scale} 
\end{equation}
where $g_*$ is the number of relativistic degrees of freedom at temperature T. In the radiation dominated epoch, using Eq.(\ref{formation-time}) and Eq.(\ref{temperature-scale}) the horizon mass can be expressed as
\begin{equation}
M_{\rm BH} (R) =
{4 \pi \over 3} \left({8 \pi G \over 3}\right)^{-1}
\left[ {H_0^2 \Omega_{\rm M} \over 1+z_{\rm eq}} \left({g_{* \rm eq} 
\over g_*}\right)^{1/3}\right]^{1/2} 
R^2\,,
%\nonumber \\
%&=&
%10^{15}  \left( g_* \over 100 \right)^{-1/6}
%\left( R\over 6.2 \times 10^8\ {\rm cm}\right)^2 ~{\rm g}\,. 
\label{bh-mass}
\end{equation}
where $g_{* \rm eq}$ and $z_{\rm eq}$ denote the number of relativistic degrees of freedom and the redshift of the matter-radiation equality, respectively.

Tashiro et al.\cite{Tashiro:2008sf} have calculated the number density of PBHs in the mass interval $M_{\rm BH}$ and $M_{\rm BH}+dM_{\rm BH}$ at any epoch, $dn(\nu, M_{\rm BH})$,

\begin{widetext}
\begin{align}
\label{pbh-numberdensity}
dn(\nu, M_{\rm BH})&= {1 \over 4 \pi ^2 M_{\rm BH}}  \left(X (n_s-1) \over 6 M_{\rm BH} \right)^{3/2}\exp \left(-{\nu^2 (n_s,M_{\rm BH})\over 2}\right)
\times \Big[{(n_s-1) \over 4 }\nu^4 (n_s,M_{\rm BH}-{3 \over 2} (n_s-3) \nu^2 (n_s,M_{\rm BH})-3\Big]dM_{\rm BH}
\nonumber\\
&\approx{1 \over 4 \pi ^2 M_{\rm BH}}  \left(X (n_s-1) \over 6 M_{\rm BH} \right)^{3/2}   
{(n_s-1)\over 4 }\nu^4 (n_s,M _{\rm BH})
\exp \left(-{\nu^2 (n_s,M_{\rm BH}) \over 2}\right)dM_{\rm BH}\,.
\end{align}
\end{widetext}
Here,
\begin{equation}
\nu (n_s,M_{\rm BH}) = \left[ 2 \left(k_0 ^2 M_{\rm BH} / X  \right)^{(n_s-1)/2} 
\over {\cal R}_c \Gamma \left((n_s-1)/2 \right) \right]^{1/2} \zeta _{\rm th}\,.
\end{equation}
We have considered the high peak limit, $\nu \gg 1$ in obtaining Eq.(\ref{pbh-numberdensity}).

\section{Estimation of injected energy via PBH\MakeLowercase{s} radiation}\label{estimate-energy}

In this section, we summarize the calculations for the estimation of the energy $Q$ pumped into the background fluid due to the Hawking radiation of PBHs in the redshift range $z_{\mu}$ and $z_{i}$. Consider a Schwarzschild black hole with mass $M_{\rm BH}$ which emits particles near the horizon with spin $s$ \cite{Page-1976}. The rate of total energy emitted between $E$ and $E+dE$ per degree of freedom is given by \cite{Hawking-1975,Gibbon-1991},
\begin{equation}
\dfrac{dN_{emit}}{dtdE}dE = \dfrac{\Gamma_{s}}{2\pi\hbar}\left[\exp\left(\dfrac{E}{kT(M_{\rm BH})}\right)-(-1)^{2s}\right]^{-1} dE,
\end{equation}
where the temperature of a black hole is related to its mass $M_{\rm BH}$ as
\begin{equation}
T(M_{\rm BH}) = \dfrac{1}{8\pi GM_{\rm BH}} \approx 1.0 \left(\dfrac{M_{\rm BH}}{10^{13}g}\right)^{-1} GeV.
\end{equation}
Here $\Gamma_{s}$ signifies the dimensionless absorption probability of emitted species. In case of photons, $\Gamma_s$ can be expressed as \cite{Gibbon-Webber}
\begin{equation}
\Gamma_s = \begin{cases}
    64 G^{4}M_{\rm BH}^{4}E^{4}/3 & E \ll kT(M_{\rm BH})\,,\\
    27 G^{2}M_{\rm BH}^{2}E^{2} & E \gg kT(M_{\rm BH})\,.
\end{cases}             
\end{equation}
The PBH loses its mass $M_{\rm BH}$ by virtue of Hawking evaporation. The rate of mass loss of a PBH can be given as 
\begin{equation}\label{masslosspbh}
\dfrac{dM}{dt} =  - 5.3\times 10^{25}f(M)\left(\dfrac{M}{g}\right)^{-2} \text{g $sec^{-1}$},
\end{equation}

%\begin{align}
%f(M) & = 1.569 + 0.569\Big[\exp\left(\dfrac{-0.0234}{T(M)}\right) + 6 \exp\left(\dfrac{-0.066}{T(M)}\right)\nonumber\\
% & \quad +  3 \exp\left(\dfrac{-0.11}{T(M)}\right) + \exp\left(\dfrac{-0.394}{T(M)}\right)\nonumber\\
% & \quad +  3 \exp\left(\dfrac{-0.413}{T(M)}\right) + 3 \exp\left(\dfrac{-1.17}{T(M)}\right)\nonumber\\
% & \quad +  3 \exp\left(\dfrac{-22}{T(M)}\right)\Big] + 0.963 \exp\left(\dfrac{-0.10}{T(M)}\right).
%\end{align}  

We can approximately integrate Eq.(\ref{masslosspbh}) and evaluate the time evolution of the PBH mass which was formed at some earlier epoch with an initial mass $M_{\rm BH}$ by
\begin{equation}
M(M_{\rm BH},t) \approx \left[M_{\rm BH}^{3} - 1.5\times 10^{26} f(M)t\right]^{1/3} \,,
\end{equation}
where $f(M)$ being a weak function of $M$ takes typical values viz. 1.0, 1.6, 9.8 or 13.6 for $M \gg 10^{17}$ g, $M \gg 10^{15}$ g, $M \gg 10^{13}$ g, $M \gg 10^{11}$ g respectively \cite{MacGibbon:1991tj}. We can express
the energy injection rate due to evaporating PBHs as 
\begin{equation}\label{enerinjrate}
\dot{Q}(t) = \int_{M_{min}(t)}^{M_{H}(t)} dM_{\rm BH} \frac{dn}{dM_{\rm BH}} \int_{0}^{\infty} dE a^{-3}(t)E \dfrac{dN_{emit}}{dtdE}\,,
\end{equation}
where $M_{H}(t)$ is the horizon mass at a time $t$ and $M_{min}(t)$ is the minimum initial mass of the PBHs. 
Using Eq.(\ref{energy-inj}) and Eq.(\ref{enerinjrate}), we estimate the total energy dissipated by evaporating PBHs into the background fluid between the redshifts $z_{\mu}$ and $z_{i}$.

\setlength{\textfloatsep}{5pt}
\begin{figure*}[htbp]
\epsfig{figure=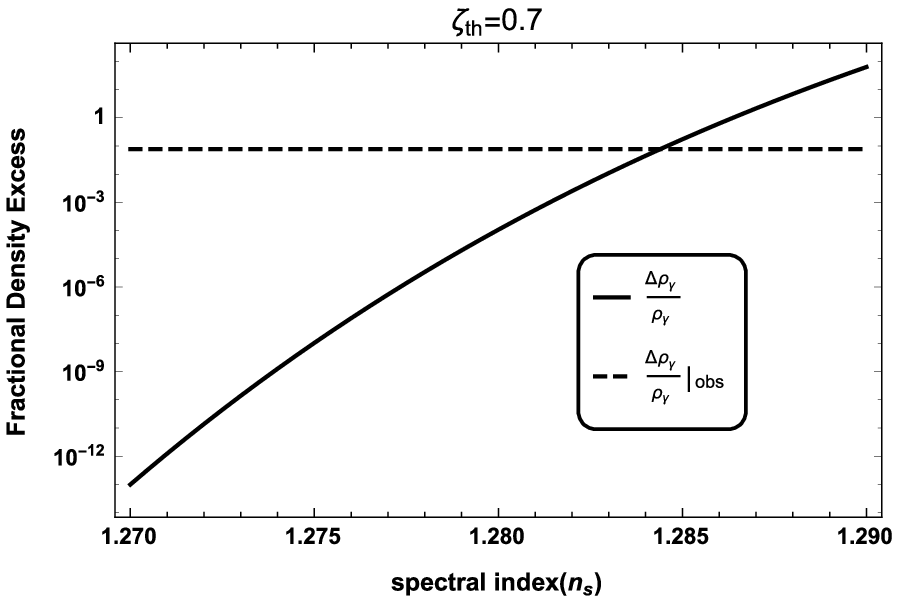,height=6cm,width=8cm,angle=0}
\epsfig{figure=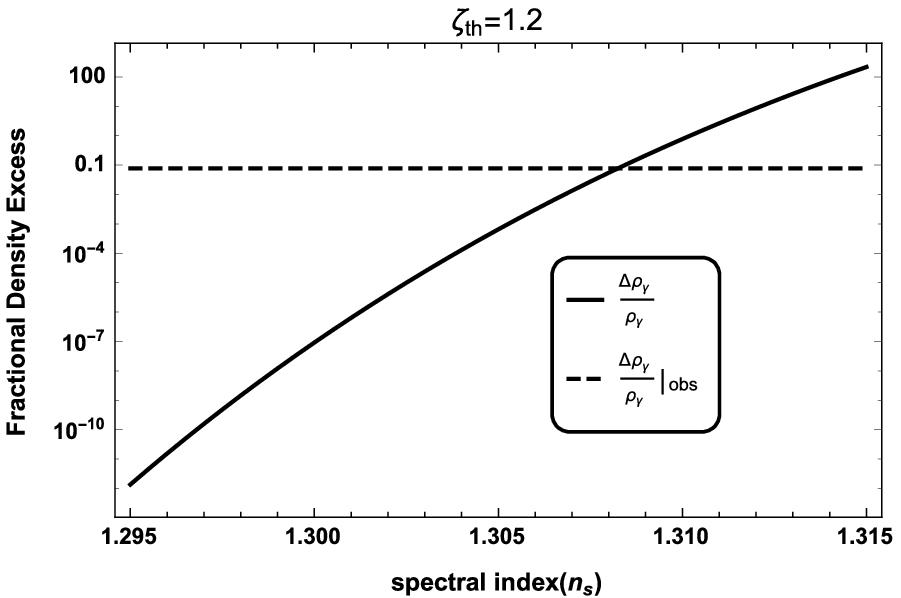,height=6cm,width=8cm,angle=0}

\caption{The fractional density excess of injected energy as a function of spectral index. The critical threshold is assumed to be $ \zeta _{\rm th} = 0.7$ (left) and $ \zeta _{\rm th}=1.2$ (right) respectively. The dashed line is the observational upper bound on the density fraction $ \dfrac{\Delta \rho_{\gamma}}{\rho_{\gamma}}$ and the region below this is allowed. The obtained upper limit on the spectral index is $ n_s < 1.284 $ for $ \zeta _{\rm th} = 0.7 $ and $ n_s < 1.308 $ for $ \zeta _{\rm th} = 1.2 $.}
\label{fig1}
\end{figure*}

\setlength{\textfloatsep}{5pt}
\begin{figure*}[htbp]
\epsfig{figure=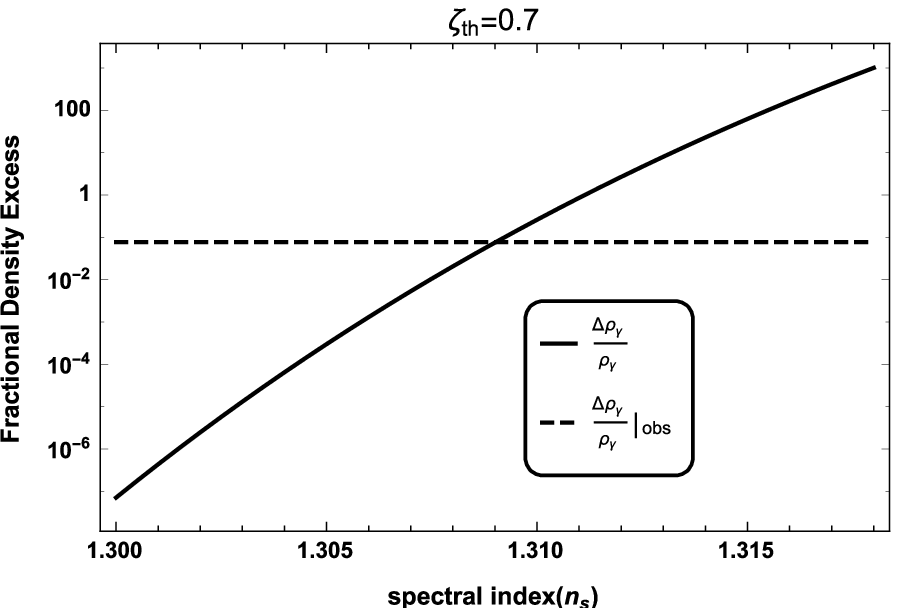,height=6cm,width=8cm,angle=0}
\epsfig{figure=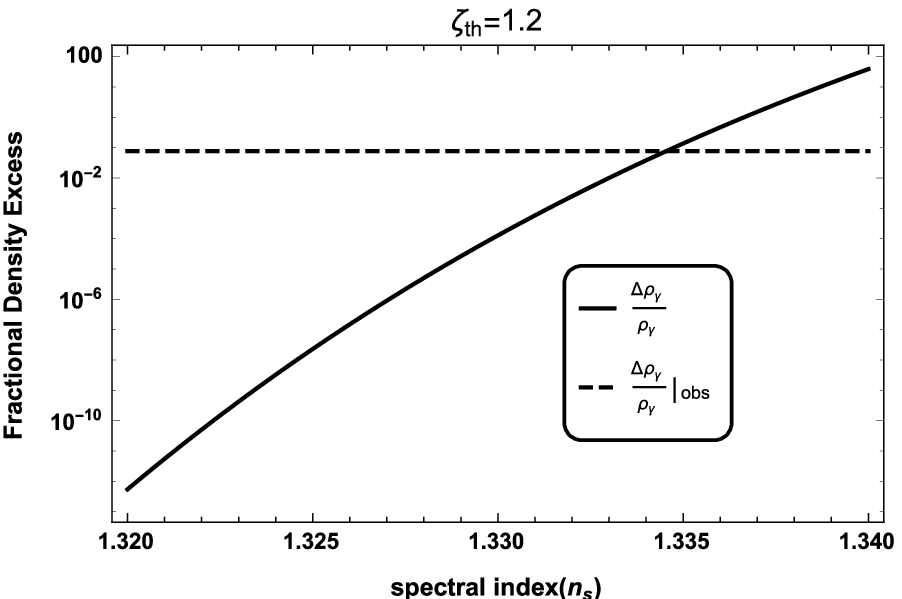,height=6cm,width=8cm,angle=0}

\caption{The fractional density excess of injected energy as a function of spectral index. The critical threshold is assumed to be $ \zeta _{\rm th} = 0.7$ (left) and $ \zeta _{\rm th} = 1.2$ (right). The dashed line is the observational upper bound on the density fraction $ \dfrac{\Delta \rho_{\gamma}}{\rho_{\gamma}}$ and the region below this is allowed. The obtained upper limit on the spectral index is $ n_s < 1.309 $ for $ \zeta _{\rm th} = 0.7 $ and $ n_s < 1.334 $ for $ \zeta _{\rm th} = 1.2 $.}
\label{fig2}
\end{figure*}
\section{The PBH Abundance}\label{pbh-abund}
To describe the PBH mass abundance, we define a function $\beta (M_{\rm BH,0})$ which represents the fraction of total energy density which collapses to form the black holes, as
\begin{align*}
\beta(M_{\rm BH,0})&\equiv \frac{\rho_{\rm BH}(M_{\rm BH,0})}{\rho}=\frac{\int_{M_{\rm BH,0}}^{\infty}M_{\rm BH}~ d n(\nu,M_{\rm BH})}{\rho}\\
\end{align*}
Here $\rho_{\rm BH}$ denotes the energy density of black holes and $\rho$ represents the background energy density of the universe at the formation of black hole of mass $M_{\rm BH,0}$.
% Using Eq. \eqref{density-peak}, we get
%\begin{align}
%\beta(M_{\rm BH})&={1 \over (2 \pi)^3} {(n_{s}-1)^{3/2} \over 4 \sqrt{6}} {(\nu^2 -1)}
%\exp\left(-{\nu^2 \over 2}\right).
%\end{align}

\section{Results and Discussion}\label{constraints}
We assumed power-law primordial spectrum, ${\cal P}_{{\cal R}} ={{\calR}_c} (k/k_0)^{n_s-1}$. The mass range of the PBHs constrained in this study lies between $10^{9}$ and $10^{11}$g, which corresponds to those PBHs whose maximal evaporation took place in the redshift range $10^{6} < z < 10^{9}$. We have obtained an upper bound on the spectral index of density fluctuations by using observational parameters from WMAP as well as Planck. In Fig.\ref{fig1} we have shown the results for the case of WMAP where ${{\cal R}_c}$ is $(24.0 \pm 1.2) \times 10^{-10}$ at the scale $k_0 =0.002 {\rm Mpc}^{-1}$ \cite{Spergel:2006hy}. In this case, we found the upper bound on the spectral index to be $ n_s < 1.284 $ for $ \zeta _{\rm th} =0.7 $ and $ n_s < 1.308 $ for $ \zeta _{\rm th} = 1.2 $. On the other hand, in Fig.\ref{fig2} using Planck results for cosmological parameters where ${{\cal R}_c}$ is $(21.37 \pm 1.2) \times 10^{-10}$ at the scale $k_0 =0.05 {\rm Mpc}^{-1}$ \cite{Ade:2015xua} we have found the upper bound on the spectral index to be $ n_s < 1.309 $ for $ \zeta _{\rm th} =0.7 $ and $ n_s < 1.334 $ for $ \zeta _{\rm th} = 1.2 $. In obtaining the above mentioned bounds on $n_s$ we have only considered the contribution of photons that are being directly emitted from the black holes. However, if the contribution of secondary photon emission is considered, our bounds on $n_s$ remains same upto the second decimal place.

In our work, we have considered a blue spectrum for the initial density fluctuations. To normalise the spectrum, we have used WMAP and Planck data. The pivot scale for normalisation considered in the Planck data ($k=0.05$ Mpc$^{-1}$) is greater than that of WMAP data ($k=0.002$ Mpc$^{-1}$) but the amplitudes on both the scales are almost same. Since we have considered a blue spectrum, it turns out that the spectrum amplitude for the scales of our interest for Planck data is smaller compared to the WMAP data. Due to this fact, the number of black holes formed for the scales of our interest are larger for WMAP data than the Planck data. This is the reason, we are getting a tighter constraint on $n_s$ in the case of WMAP data compared to the case of Planck data.  

\setlength{\textfloatsep}{5pt}
\begin{figure}[t]
\epsfig{figure=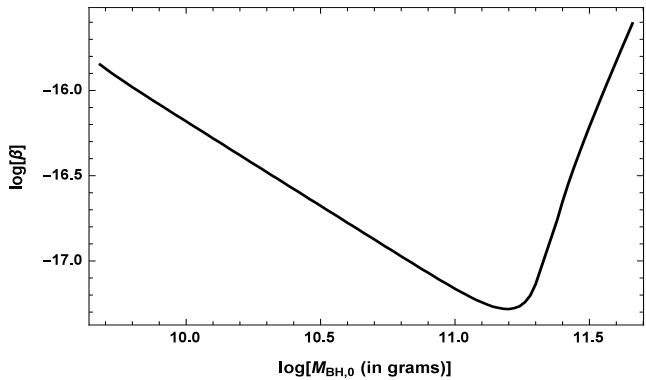,height=5cm,width=8.5cm,angle=0}
\caption{In this figure, we have shown the upper bound on PBH abundance as a function of minimum mass of the black hole using the constraint on the fractional density excess dissipated to the background fluid. Here we have used Planck data and $\zeta_{th}=0.7$.}
\label{beta_function}
\end{figure}

In estimating the fractional density excess for power law behaviour with different values of $n_s$, we assume that $M_{BBN}$ mass black hole formed after reheating. Hence reheating takes place at a temperature larger than the temperature of the universe when the black holes of mass $M_{BBN}$ is formed ($M_{BBN}$ is the mass of that black hole which evaporate away at the epoch of BBN). If the reheating takes place below this temperature then the masses of the black holes formed will be greater than $M_{BBN}$. It means that as the temperature of reheating decreases, mass range of the black holes which evaporate away between the epoch of BBN and CMB-$\mu$ distortions also decreases. The maximum contribution to the excess density fraction comes from the minimum mass of the black hole in this range. This allows us to constrain the mass fraction of the black holes using the constraint on the spectral index. Further we take different reheating temperatures and calculate their corresponding mass scales and determine the bounds on the spectral index. Using this mass and the corresponding bound on spectral index, we calculate the beta function. 

In Fig.(\ref{beta_function}) we show the upper bound obtained on the PBHs abundance as a function of their masses. The upper bound turns out to be $\beta < 10^{-17}$ for the investigated mass range ($10^9 ~g~ \text{to}~ 10^{11} ~g$).

\section{Conclusion}\label{concl}
In the present work, we have considered evaporating PBHs that had injected energy into the background fluid. Since the processes like double Compton scattering and Bremsstrahlung are very significant during the early phases of the
universe, the number density of background photons increases while the number density of baryons remain unchanged. As a result of this, the baryon-photon number ratio decreases due to additional
reheating. Since modern cosmological observations such as BBN and CMB well constraints this baryon-
photon number ratio we can use this to obtain an upper bound on the spectral index of density fluctuations at small scales which are difficult to be probed through the observations of CMB anisotropies and distortions.
Using Planck results for cosmological parameters we obtained the upper bound on $ n < 1.309 $ for $ \zeta _{\rm th} =0.7 $ and $ n < 1.334 $ for $ \zeta _{\rm th} = 1.2 $ respectively. In our study, the constraint on the PBH density fraction is found to be $\beta < 10^{-17}$. This is a significant probe of the amplitude of density fluctuations on small scales that cannot be accessed by CMB anisotropies.

\begin{acknowledgements}

GG would like to thank University Grants Commission, Govt. of India to provide financial support via Dr. D. S. Kothari Postdoctoral Fellowship. RS acknowledges CSIR, India for the financial support through grant 09/045(1343)/2014-EMR-I. TRS acknowledges SERB for the project grant EMR/2016/002286. The authors also acknowledge the services of IRC, Delhi. 

\end{acknowledgements}

\end{document}